\documentclass[preprint,english,showpacs,11pt,floatfix]{revtex4}

\usepackage{babel,amsmath,amssymb,dcolumn}
\usepackage[dvips]{graphics}

\begin{document}

\title{Thermodynamic Geometry and Critical Behavior of Black Holes}

\author{Jianyong Shen$^1$, Rong-Gen Cai\footnote{Email address: cairg@itp.ac.cn}$^{,2}$,
 Bin Wang\footnote{Email address: wangb@fudan.edu.cn}$^{,1}$ and Ru-Keng Su\footnote{Email address:
 rksu@fudan.ac.cn}$^{,3,1}$}
 \affiliation{$^1$Department of Physics,
Fudan University, Shanghai 200433,  China \\
 $^2$Institute of Theoretical
Physics, Chinese Academy of Sciences, \\
P.O. Box 2735, Beijing 100080, China\\
$^3$ China Center of Advanced Science and Technology (World
Laboratory) P.O. Box 8730, Beijing 100080,  China }

\begin{abstract}
Based on the observations that there exists an analogy between the
Reissner-Nordstr\"om-anti-de Sitter (RN-AdS) black holes and the
van der Waals-Maxwell liquid-gas system, in which a correspondence
of variables is $(\phi, q) \leftrightarrow (V,P)$, we study the
Ruppeiner geometry, defined as Hessian matrix of black hole
entropy with respect to the internal energy (not the mass) of
black hole and electric potential (angular velocity), for the RN,
Kerr and RN-AdS black holes. It is found that the geometry is
curved and the scalar curvature goes to negative infinity at the
Davies' phase transition point for the RN and Kerr black holes.
 Our result for the RN-AdS black holes is also in good agreement
with the one about phase transition and its critical behavior in
the literature.

\end{abstract}


\maketitle

\section{Introduction}

One of important characteristics of a black hole is its
thermodynamic property: a black hole has Hawking temperature
proportional to its surface gravity on the horizon of the black
hole, entropy proportional to its horizon area~\cite{Haw,Bek}, and
they satisfy the first law of black hole thermodynamics
~\cite{firstlaw}, although the statistical origin of the black
hole entropy still remains obscure. In general relativity, the
most general stationary black hole solution with asymptotical
flatness is the Kerr-Newman solution, which describes a rotating,
charged black hole with only three parameters: mass, electric
charge and angular momentum. This characteristic is called no hair
theorem of black holes. For the Schwarzschild black hole (static,
spherically symmetric black hole without electric charge), its
Hawking temperature is inversely proportional to its mass, the
heat capacity of the black hole is therefore always negative and
the black hole is thermodynamically unstable. However, for the
Reissner-Nordstr\"om (RN) black hole (static, spherically
symmetric black hole with electric charge), Kerr black hole
(rotating black hole without electric charge), and more general
Kerr-Newman black hole, their heat capacity is positive in some
parameter region and negative in other region, and between them,
the heat capacity diverges. As one knows, the divergence of heat
capacity is the indication of a second order phase transition in
the ordinary thermodynamic systems. Just on the basis of the
divergence of heat capacity, Davies~\cite{Davies1} argued that
phase transition appears in black hole thermodynamics and the
phase transition point is the one where the heat capacity diverges
(see also \cite{Hut}). Some authors investigated different aspects
of this critical point~\cite{list1} and found that some critical
exponents related to this critical point obey the scaling laws. On
the other hand, some people argued that there exists a critical
point at the extremal limit of black holes and a second order
phase transition takes place from an extremal black hole to its
non-extremal counterpart, some critical exponents related to this
critical point also satisfy scaling laws~\cite{list2}.

Over the past decade, due to the AdS/CFT correspondence (for a
review see \cite{s1}), there has been a lot of interest in the
thermodynamics of various black holes in anti-de Sitter(AdS)
space. It was convincingly argued by Witten~\cite{witt} that the
thermodynamics of black holes in AdS space can be identified with
that of the dual conformal field theory (CFT) residing on the
boundary of the AdS space. Therefore by studying thermodynamics
and phase structure of black holes in AdS space, one can gain sone
insights into corresponding ones of dual strong coupling CFTs and
{\it vice versa}. For instance, the Hawking-Page phase
transition~\cite{HP} between the large stable Schwarzschild black
hole phase and thermal AdS space phase can be explained as the
deconfinement/confinement phase transition in the dual gauge field
theory.  In addition, Chamblin {\it et al.}~\cite{Cham1,Cham2}
investigated thermodynamics of RN  black holes in AdS space, and
revealed rich phase structure of RN-AdS black holes in the fixed
charge ensemble.  In the spirit of AdS/CFT correspondence, the
charge carried by the RN-AdS black hole can be explained as the
background current to which the dual CFT couples. The phase
structure of RN-AdS black holes is in good agreement with
expectation from the dual CFT.  In particular, we would like to
stress here that the point where the heat capacity diverges with
fixed charge for the RN-AdS black hole is indeed a critical point
of a second order phase transition.  Obviously, the results given
by Chamblin {\it et al.} also supports the viewpoint of Davies
that the divergence point of heat capacity of black holes is a
phase transition point, since according to the holographic
principle that a theory with gravity can be dual to a theory
without gravity in one dimension fewer~\cite{holo}, the
thermodynamics of black holes in asymptotically flat spacetime can
also be identified with that of a dual theory without gravity,
although we do not know so far what the dual theory is.

On the other hand, one can introduce some standard geometrical
ideas in ordinary thermodynamics (for a review see \cite{Rupp1}).
Weinhold~\cite{Wein} was the first to introduce the geometrical
concept into the thermodynamics, he suggested a sort of Riemannian
metric defined as the second derivatives of internal energy $U$
with respect to entropy and other extensive quantities of a
thermodynamic system. However, the geometry based on this metric
seems  physically meaningless in the context of purely equilibrium
thermodynamics. Later soon, Ruppeiner~\cite{Rupp2} introduced a
metric, defined as the second derivatives of entropy $S$ with
respect to the internal energy and other extensive quantities of a
thermodynamic system. It turns out that the Ruppeiner metric is
conformally related to the Weinhold metric with the inverse
temperature as the conformal factor. The Ruppeiner geometry has
its physically meanings in the fluctuation theory of equilibrium
thermodynamics. The components of the inverse Ruppeiner metric
gives second moments of fluctuations. Since the proposal of
Ruppeiner, many investigations have been carried out on the
physically meanings of the Ruppeiner geometry in various
thermodynamic systems such as the ideal classical gas,
multicomponent ideal gas, ideal quantum gas, one-dimensional Ising
model, van der Waals model and so on. In particular, it was found
that the Ruppeiner geometry carry information of phase structure
of thermodynamic system; and scalar curvature of the metric
diverges (goes to negative infinity) at the phase transition and
critical point, which shows a strong correlation of system. For
thermodynamic systems with no statistical mechanical interactions
(for example, ideal gas), the scalar curvature is zero and the
Ruppeiner metric is flat. These have been summarized in the review
paper~\cite{Rupp1} (some recent works on the Ruppeiner geometry
see~\cite{John}).

Since the Ruppeiner geometry in some sense can reveal some
features of statistical mechanical models, it is therefore of
great interest to apply the Ruppeiner geometry to black hole
thermodynamics, because so far we do not know their statistical
models behind thermodynamics of most black holes, except for a few
black holes in superstring/M theories. Indeed, the geometric
approach to thermodynamics was first to be introduced to the black
hole thermodynamics and to discuss the critical behavior in moduli
space by Ferrara {\it et al.}~\cite{Ferr}. In particular, it was
found that the Weinhold metric is proportional to the metric on
the moduli space for supersymmetric extremal black holes, whose
Hawking temperature is zero, and the Ruppeiner metric governing
fluctuations naively diverges, which is consistent with the
argument that near extreme the thermodynamic description  breaks
down~\cite{Preskill}. The geometric approach has also been used to
the BTZ black hole~\cite{Caicho} and RN black holes and Kerr black
holes in diverse dimensions~\cite{Aman1,Aman2}. The authors of
\cite{Aman1,Aman2} found that considering the entropy as a
function of the mass and charge (angular momentum)of RN (Kerr)
black hole, the Ruppeiner metric is always flat (curved) for the
RN (Kerr) black holes and the scalar curvature vanishes for the RN
black hole, while it diverges at the extremal Kerr black hole.
Namely, the Ruppeiner metric for the RN black holes is quite
different from that of the Kerr black holes. In addition there is
no special occurring at the phase transition point of Davies.

Based on the observations made by Chamblin~{\it et
al.}~\cite{Cham1,Cham2} and by Wu~\cite{wu} for the phase
structure of RN-AdS black holes, in the present paper we will
reconsider the thermodynamic geometry of RN, Kerr and RN-AdS black
holes in four dimensions. Generalization to other higher
dimensions is straightforward. One of important observations in
\cite{Cham1,Cham2} is that the phase structure of RN-AdS black
holes is analogous to that of the van der Waals-Maxwell liquid-gas
system, maybe it is beyond just an analogy.  The phase diagram
($Q-T$ diagram, where $Q$ and $T$ denote the charge and Hawking
temperature ) of the RN-AdS black holes is quite similar to the
phase diagram ($P-T$ diagram, where $P$ and $T$ stand for the
pressure and temperature ) of the van-der Waals-Maxwell system;
the equation of state $Q-\phi$ diagram (where $\phi$ denotes the
electric potential on the horizon of black hole) of the RN-AdS
black holes is similar to that of $P-V$ diagram ($V$ is the
volume) of the van der Waals-Maxwell system. That is, in this
analogy of the RN-AdS black hole to the van der Waals-Maxwell
system, the electric potential $\phi$ plays the role of the
extensive quantity volume $V$ and the charge $Q$ the role of the
intensive quantity pressure $P$. In addition, let us notice that
an appropriate order parameter is the difference of the electric
potentials in two phases in the RN-AdS black holes, while the
order parameter is the difference of energy densities between
liquid phase and gas phase in the van der Walls-Maxwell system.
Furthermore, we redefine the internal energy $U$ of black holes (
the ADM gravitational mass of black holes subtracted by the energy
of electric field outside the black hole horizon (for charged
black holes) and/or the energy due to the rotation (for the
rotating black holes). Considering the entropy of black holes as a
function of the internal energy and electric potential (angular
velocity) of RN, Kerr and RN-AdS black holes, we obtain the
Ruppeiner metric and calculate the scalar curvature of geometry.
We find that the result is consistent with the expectation: the
scalar curvature diverges at the phase transition points of Davies
for the RN and Kerr black holes; the case of the RN-AdS black
holes is also in complete agreement with \cite{Cham1,Cham2,wu}.

This paper is organized as follows. In Sec.~II and Sec.~III, we
will respectively discuss the thermodynamic geometry of the RN
black hole and Kerr black hole, and calculate their scalar
curvature $\hat R$, which indeed goes to negative infinity at the
phase transition point of Davies. In Sec.~IV, we will first
briefly review the thermodynamics of the RN-AdS black hole and
then study its thermodynamic geometry. We end this paper with
conclusion and discussions in Sec.~V.

\section{Thermodynamic Geometry of RN Black Holes}

The charged static black hole is known as the RN black hole whose
metric  is  given by
\begin{equation} \label{e1}
ds^2  =  - f(r)dt^2  + f(r)^{ - 1} dr^2  + r^2 d\Omega ^2,
\end{equation}
in four  dimensional spacetime, where $d\Omega ^2$ is the line
element on a unit 2-sphere and
\begin{equation} \label{e2}
f(r) = 1 - \frac{\mu }{r} + \frac{{q^2 }}{{r^2 }}.
\end{equation}
$\mu/2=M$ and $q$ is the ADM mass and electric charge of the black
hole respectively in units ($ G = c = \hbar  = k_B = 1 $) which we
use throughout this paper. The two horizons of the RN black hole,
inner Cauchy horizon located at $r_-$ and outer event horizon at
$r_+$, can be expressed by mass and electric charge
\begin{equation} \label{e3}
\mu = r_- + r_+, \quad q^2 = r_- r_+
\end{equation}
with the condition $q^2  \le \mu ^2 /4$ ruling out the naked
singularity at $r=0$. When $r_-$ equals to $r_+$, the black hole is
extremal. The entropy of the black hole is obtained by the area law
\begin{equation} \label{e4}
S=A/4=\pi r_+^2.
\end{equation}
The energy conservation law of the black hole
\begin{equation}
\label{RN1}
 dM= TdS + \phi dq
 \end{equation}
  implies the Hawking temperature $T$ and electric
potential on the event of horizon $\phi$ of the RN black hole
\begin{equation} \label{e5}
T = \left( {\frac{{\partial M }}{{\partial S}}} \right)_q  =
\frac{{r_ +   - r_ -  }}{{4\pi r_ + ^2 }},
\end{equation}
\begin{equation} \label{e6}
\phi  = \left( {\frac{{\partial M }}{{\partial q}}} \right)_S  =
\frac{{\sqrt {r_ +  r_ -  } }}{{r_ +  }} = \frac{{q}}{{r_ +  }}.
\end{equation}
The combination of Eq.(\ref{e3}), Eq.(\ref{e5}) and Eq.(\ref{e6})
provides an equation of state (EOS) of the RN black hole
$\phi=\phi(q,T)$.

Before proceeding, let us discuss in some detail what roles play
by $\phi$ and $q$ in the analogy of the RN-AdS black hole to the
van der Waals-Maxwell system, for the latter, the first law of
thermodynamics is
\begin{equation}
\label{vander}
 du=TdS -PdV,
 \end{equation}
 where $u$ is the internal energy, $P$ and $V$ are respectively
 the pressure and volume of the system. As presented in the Sec.~IV
 (also see \cite{Cham1,Cham2,wu}), although the electrical charge $q$
looks like an extensive variable and the potential $\phi$ like an
intensive one, from the isotherms in the $q - \phi$ plane (see
Fig.~\ref{f1}) and the $q-T$ phase diagram of the RN-AdS black
hole, we can see that $\phi$ and $q$, respectively, play the roles
of volume $V$ and pressure $P$ in the corresponding diagrams of
the van der Waals-Maxwell system. That is, the correspondence is
$(\phi, q) \to (V, P)$ for establishing the phase structure of the
RN-AdS black holes.  The internal energy represents the basic and
intrinsic properties of the thermodynamic system, which excludes
the contribution of the external work. For the existence of the RN
black hole, the intrinsic property is obviously determined by the
structure of the spacetime itself excluding the effects of the
static electricity. Therefore the appropriate internal energy
should be understandably written as
\begin{equation} \label{e7}
u = M  - \phi q.
\end{equation}
Expressed by the internal energy, the first law of thermodynamics
is written as
\begin{equation} \label{e8}
du = TdS  - q d\phi.
\end{equation}
Comparing (\ref{vander}) and (\ref{e8}), we can clearly establish
the correspondence $(\phi, q) \to (V, P)$. Other pieces of
 supporting evidence to express the internal energy of charged
black hole as (\ref{e7}) come as follows. When expressing the
entropy of the RN-AdS black holes in terms of the Cardy-Verlinde
formula, we noticed that the energy of electric field outside the
black hole horizon has to be subtracted from the mass $M$ of the
black hole~\cite{Cai}; For a rotating body, its internal energy is
the difference between the total energy of the body and its
kinetic energy of rotation~\cite{Lan}. It is therefore reasonable
to argue that the discussion above for the RN-AdS black hole also
holds for the asymptotically flat RN and Kerr black holes. That
is, the internal energy of the RN black hole should be defined as
in (\ref{e7}). As a state function, entropy should be a function
of internal energy.

 We now turn to the thermodynamic geometry of the RN black hole.
 The thermodynamic metric introduced in Ruppeiner's
theory \cite{Rupp1} is defined by the second derivatives of the
entropy. It is worth paying attention that the entropy inducing
the metric must be in the strict form of the function regarded as
the internal energy and the extensive variables in ordinary
thermodynamic systems
\begin{equation} \label{e8.5}
\hat g_{ab}  = \frac{{\partial ^2 }}{{\partial x^a \partial x^b
}}S(x)\quad (a,b = 1,2,...,n),
\end{equation}
where $x=(u, x^1,x^2,...,x^{n-1})$ denotes  the internal energy
$u$ and the extensive variables $x^a \quad (a \ne 1)$.  Based on
the discussions above, for the RN black hole, the thermodynamic
metric can be written down as
\begin{equation} \label{e9}
\hat g_{ab}  = \frac{{\partial ^2 }}{{\partial x^a \partial x^b
}}S(u,\phi) \quad (a,b = 1,2),
\end{equation}
where $x^1 = u$ and $x^2 = \phi$. Note that for the van der Waals
model, the entropy is a function of the internal energy $u$ and
fluid density $\rho$~\cite{Rupp1}. Using Eq.(\ref{e4}), the direct
calculation yields
\begin{equation} \label{e9.5}
\hat g = \left( {\begin{array}{*{20}c}
   { - \frac{{4\pi r_ +  ^2 \sqrt {r_ +  r_ -  } }}{{r_ +  (r_ +   - r_ -  )^2 }}}
   & {\frac{{2\pi r_ +  ^2 }}{{(r_ +   - r_ -  )^2 }}}  \\
   {\frac{{2\pi r_ +  ^2 }}{{(r_ +   - r_ -  )^2 }}}
   & {\frac{{\pi r_ +  ^3 (r_ +   + 5r_ -  )}}{{(r_ +   - r_ -  )^2 }}}  \\
\end{array}} \right)
\end{equation}
and the scalar curvature is
\begin{equation} \label{e10}
\hat R = \hat g_{ab} \hat R^{ab}  =  - \frac{{r_ +   - r_ -  }}{{\pi
r_ +  (3r_ -   - r_ +  )^2 }}.
\end{equation}
We see in our setup that the scalar curvature vanishes only at the
extremal limit where $r_+=r_-$. In a general case, the scalar
curvature does not vanish and it goes to negative infinity when
$r_+=3r_-$, which stands for a kind of phase transition or long
range correlation of the system according to the Ruppeiner's
theory~\cite{Rupp1}.  It is interesting to note that the
divergence point of the scalar curvature is just the phase
transition point of Davies~\cite{Davies1}. It is easy to check
this by calculating the heat capacity with a fixed charge
\begin{equation} \label{e11}
C_q  = T\left( {\frac{{\partial S}}{{\partial T}}} \right)_q  =
\frac{{2\pi r_ + ^2 (r_ +   - r_ -  )}}{{3r_ -   - r_ +  }},
\end{equation}
which is singular at $3r_ - = r_ +$ and indicates that the black
hole has a second order phase transition \cite{Davies1,list1}.
Therefore, the Ruppeiner's theory well describes the critical
behavior of RN black hole thermodynamics as it does in ordinary
thermodynamic systems, after carefully understanding of some
quantities of black hole thermodynamics.

\section{Thermodynamic Geometry of Kerr Black Holes}

The rotating black hole without charge is known as Kerr black
hole, whose line element is
\begin{eqnarray} \label{e12}
ds^2  &=&  - \frac{{\Delta  - a^2 \sin ^2 \theta }}{\Sigma }dt^2  -
\frac{{2a\sin ^2 \theta (r^2  + a^2  - \Delta )}}{\Sigma }dtd\phi  +
\frac{{(r^2  + a^2 )^2  - \Delta a^2 \sin ^2 \theta }}{\Sigma }\sin
^2 \theta d\phi ^2 \nonumber \\
&& + \frac{\Sigma }{\Delta }dr^2  + \Sigma d\theta ^2,
\end{eqnarray}
where
\begin{equation} \label{e13}
\Sigma  = r^2  + a^2 \cos ^2 \theta \quad \Delta  = r^2  + a^2 - \mu
r.
\end{equation}
Here $\mu/2= M$ and $J=a M=a \mu/2$ are the ADM mass and the angular
momentum of the Kerr black hole respectively. Both the outer and
inner horizons ($r_+$ and $r_-$) are given by $\Delta  = r^2  + a^2
- \mu r = 0$ and have the relation
\begin{equation} \label{e14}
\mu  = r_ +   + r_ -  \quad J = a M = \frac{1}{2}\sqrt {r_ +  r_ - }
(r_ +   + r_ -  ).
\end{equation}
By the area law, the entropy of the Kerr black hole is
\begin{equation} \label{e15}
S = \frac{1}{4}A = \pi (r_ +  ^2  + a^2 ) = \pi (r_ +  ^2  + r_ + r_
- ).
\end{equation}
According to the energy conservation law $dM  = TdS + \Omega _H
dJ$, the Hawking temperature and the angular velocity of the outer
horizon are respectively
\begin{equation} \label{e16}
T = \left( {\frac{{\partial M}}{{\partial S}}} \right)_J  =
\frac{{r_ +   - r_ -  }}{{4\pi (r_ + ^2  + r_ +  r_ -  )}}
\end{equation}
and
\begin{equation} \label{e17}
\Omega _H  = \left( {\frac{{\partial M}}{{\partial J}}} \right)_S
= \frac{{\sqrt {r_ +  r_ -  } }}{{2(r_ + ^2  + r_ +  r_ -) }}.
\end{equation}
Just like the case in the RN black hole, we are able to establish
thermodynamics of the Kerr black hole by using the correspondence
$(\Omega _H, J) \to (V, P)$. The internal energy of the Kerr black
hole should exclude the kinetic energy of rotation, that is
\begin{equation} \label{e18}
u = M  - \Omega _H J
\end{equation}
and the first law of thermodynamics can be written  down as
\begin{equation} \label{e19}
du = TdS  - J d\Omega _H.
\end{equation}
By the definition of the thermodynamic metric in Eq.(\ref{e8.5}),
for the Kerr case, we have
\begin{equation} \label{e20}
\hat g_{ab}  = \frac{{\partial ^2 }}{{\partial x^a \partial x^b
}}S(u,\Omega _H ),
\end{equation}
where $a,b = 1,2$ and $x^1=u,x^2=\Omega_H$.  A simple calculation
gives
\begin{equation} \label{e20.5}
\hat g = \left( {\begin{array}{*{20}c}
   {\frac{{2\pi (r_ +   + r_ -  )^3 }}{{(r_ +   - r_ -  )^3 }}} &
   {\frac{{8\pi r_ +  ^2 \sqrt {r_ +  r_ -  } (r_ +   + r_ -  )^2 }}{{(r_ +   - r_ -  )^3 }}}  \\
   {\frac{{8\pi r_ +  ^2 \sqrt {r_ +  r_ -  } (r_ +   + r_ -  )^2 }}{{(r_ +   - r_ -  )^3 }}} &
   {\frac{{2\pi r_ +  ^2 (r_ +   + r_ -  )^3 (r_ +  ^2  + 6r_ +  r_ -   - r_ -  ^2 )}}{{(r_ +   - r_ -  )^3 }}}  \\
\end{array}} \right).
\end{equation}
The  scalar curvature of the Kerr black hole is
\begin{equation} \label{e21}
\hat R = \hat g_{ab} \hat R^{ab}  =  - \frac{{4(r_ +   + r_ -  )(r_
+ ^2  - 9r_ - ^2 )}}{{\pi (r_ +   - r_ -  )(r_ +  ^2  - 6r_ +  r_ -
- r_ -  ^2 )^2 }}.
\end{equation}
We see that $\hat R$ naively diverges at the extreme limit of the
Kerr black hole  where $r_+ = r_-$, which is of less physically
interest since at the extremal limit the Hawking temperature
vanishes, the thermodynamics description breaks down as mentioned
above.  An interesting divergence of $\hat R$ for a non-extremal
black hole occurs at $r_ + ^2 - 6r_ + r_ - - r_ -  ^2 = 0$, at
which the heat capacity with a fixed angular momentum
\begin{equation} \label{e22}
C_J  =  - \frac{{2\pi r_ +  (r_ +   - r_ -  )(r_ +   + r_ -  )^2
}}{{r_ +  ^2  - 6r_ +  r_ -   - r_ -  ^2 }}
\end{equation}
also becomes singular and stands for thermodynamic critical
phenomena. Again this is also consistent with the phase transition
point of Kerr black hole given by Davies~\cite{Davies1}. In
addition, let us note that the scalar curvature vanishes at
$r_+=3r_-$, namely, $a=3M/4$. At the moment we do dot know whether
there are any special physical meanings by the vanishing of the
scalar curvature at this point.

\section{Thermodynamic Geometry of RN-AdS Black Holes}

\begin{figure}
\resizebox{0.5\linewidth}{!}{\includegraphics*{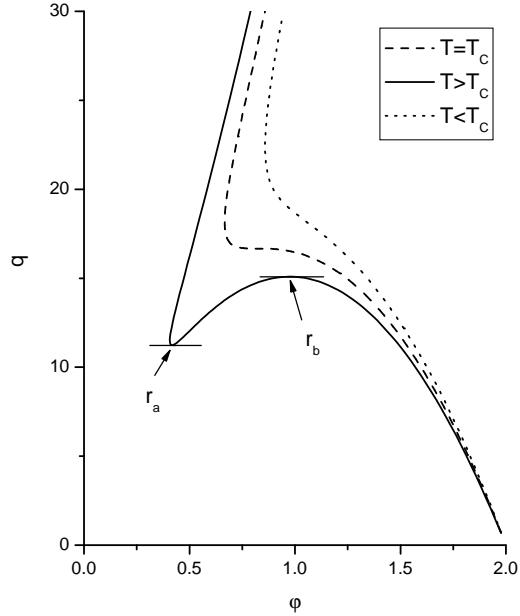}}
\caption{The EOS of the RN-AdS black hole when the temperature $T$
is equal to, less and greater than the critical temperature $T_c$
respectively. $r_a$ and $r_b$ are the values of $r_+$ when $q$
reaches the local minimum and maximum, which corresponds to the
divergences of $C_q$.}
\label{f1}
\end{figure}

Compared to the RN and Kerr black holes, the RN-AdS black holes
have much more rich thermodynamic characteristics including phase
transition and critical behavior~\cite{Cham1,Cham2,wu}, which also
have good agreements in the frame of the thermodynamic geometry.
The metric of the RN-AdS black hole in four dimensions is
\begin{equation} \label{e23}
ds^2  =  - f(r)dt^2  + f(r)^{ - 1} dr^2  + r^2 d\Omega ^2,
\end{equation}
where
\begin{equation} \label{e24}
f(r) = 1 - \frac{\mu }{r} + \frac{{q^2 }}{{r^2 }} + \frac{{r^2
}}{{l^2 }}
\end{equation}
with the negative cosmological constant $\Lambda = -3 / l^2$. The
gravitational mass $M=\mu/2$ and the electric charge $q$ are given
by the inner and outer horizons
\begin{equation} \label{e25}
\mu  = r_ +   + r_ -   + \frac{{r_ +  ^4  - r_ -  ^4 }}{{l^2 (r_ + -
r_ -  )}},
\end{equation}
\begin{equation} \label{e26}
q^2  = r_ +  r_ -  \left(1 + \frac{{r_ +  ^3  - r_ -  ^3 }}{{l^2
(r_ + - r_ -  )}}\right).
\end{equation}
The black hole entropy $S$ is
\begin{equation} \label{e27}
S = \frac{1}{4}A = \pi r_ +  ^2.
\end{equation}
Through the energy conservation law $dM=TdS +\phi dq$, we have the
Hawking temperature and electric potential on the horizon of the
black hole
\begin{equation} \label{e28}
T = \left( {\frac{{\partial M }}{{\partial S}}} \right)_q  =
\frac{{(r_ +   - r_ -  )(l^2  + 3r_ + ^2  + 2r_ +  r_ -   + r_ -
^2 )}}{{4\pi l^2 r_ + ^2 }}
\end{equation}
and
\begin{equation} \label{e29}
\phi  = \left( {\frac{{\partial M }}{{\partial q}}} \right)_S  =
\frac{{\sqrt {r_ +  r_ -  [1 + (r_ + ^2  + r_ +  r_ -   + r_ - ^2
)/l^2 ]} }}{{r_ +  }} = \frac{{q}}{{r_ +  }}.
\end{equation}
The EOS of the RN-AdS black hole $q=q(\phi,T)$ can be obtained by
using Eq.~(\ref{e25})-Eq.~(\ref{e29}) and thus the isotherm can be
drawn on the $q- \phi$ state plane as shown in Fig.\ref{f1}. We
can see indeed that the isotherms on the $q-\phi$ state plane is
quite similar to those on the $P-V$ diagram of the van der Waals
model. We note that when $T>T_c$, there are a locally maximum
point and a locally minimum point, both are determined by
\begin{equation} \label{e31}
\left( {\frac{{\partial q}}{{\partial \phi }}} \right)_T  =
\frac{{r_ +  (3r_ + ^4  - l^2 r_ +  ^2  + 3l^2 q^2 )}}{{2(3r_ + ^4
- l^2 r_ +  ^2  + l^2 q^2 )}} = 0.
\end{equation}
The isotherm can be divided into three branches in this stage as
suggested in~\cite{Cham1}.
``Branch 3" is the branch which extends all the way from $q=
\infty$ and terminates at $dq/d \phi =0$. This branch is
electrically stable and has negative free energy in the its most
part. ``Branch 2" covers the branch between the locally maximum
and minimum points, where there is an instability by thermal
fluctuations like the case of Van de Waals'
gas~\cite{Cham1,Cham2}, in spite of its electrical stability. The
rest branch is called ``Branch 1". Hence, we still can use the
``equal area law" to get the first order phase transition. The
heat capacity for a fixed charge is
\begin{eqnarray} \label{e30}
C_q  &=& T\left( {\frac{{\partial S}}{{\partial T}}} \right)_q  =
\frac{{2\pi r_ + ^2 (r_ +   - r_ -  )(l^2  + 3r_ + ^2  + 2r_ +  r_ -
+ r_ - ^2 )}}{{3r_ + ^3  - l^2 r_ +   + 3r_ -  (l^2  + r_ + ^2  + r_
+  r_ -   + r_ - ^2 )}} \nonumber \\
&=& \frac{{2\pi r_ + ^3 (r_ +   - r_ - )(l^2 + 3r_ + ^2  + 2r_ +
r_ -   + r_ - ^2 )}}{{3r_ + ^4  - l^2 r_ +^2   + 3l^2 q^2 }},
\end{eqnarray}
and it has singularities at $3r_ + ^4  - l^2 r_ +  ^2  + 3l^2 q^2
=0$, whose roots are
\begin{equation} \label{e28.5}
r_ a   = \left ( \frac{l}{6}(1 - \sqrt {1 - \frac{{36q^2 }}{{l^2
}}} )\right ) ^{1/2}, \quad r_ b   = \left ( \frac{l}{6}(1 + \sqrt
{1 - \frac{{36q^2 }}{{l^2 }}})\right ) ^{1/2},
\end{equation}
which correspond to the local maximum and minimum in Fig.\ref{f1}.
The states between these two points on the isotherm of $T>T_c$,
i.e., $r_a < r_+ <r_b$, stand on the Branch 2 and are unstable.
When $T=T_c$, the Branch 2 shrinks to one critical point
\begin{equation} \label{e32}
\left( {\frac{{\partial q}}{{\partial \phi }}} \right)_T  = \left(
{\frac{{\partial ^2 q}}{{\partial \phi ^2 }}} \right)_T  = 0,
\end{equation}
where the second order phase transition happens and $r_a = r_b$ with
$r_c^2=l^2/6,q_c^2=l^2/36$. Below the critical temperature $T_c$,
the Branch 2 vanishes and there is a transition of electrical
stability at $dq/ d\phi = \infty$ between Branch 3 and Branch 1.

\begin{figure}
\resizebox{0.5\linewidth}{!}{\includegraphics*{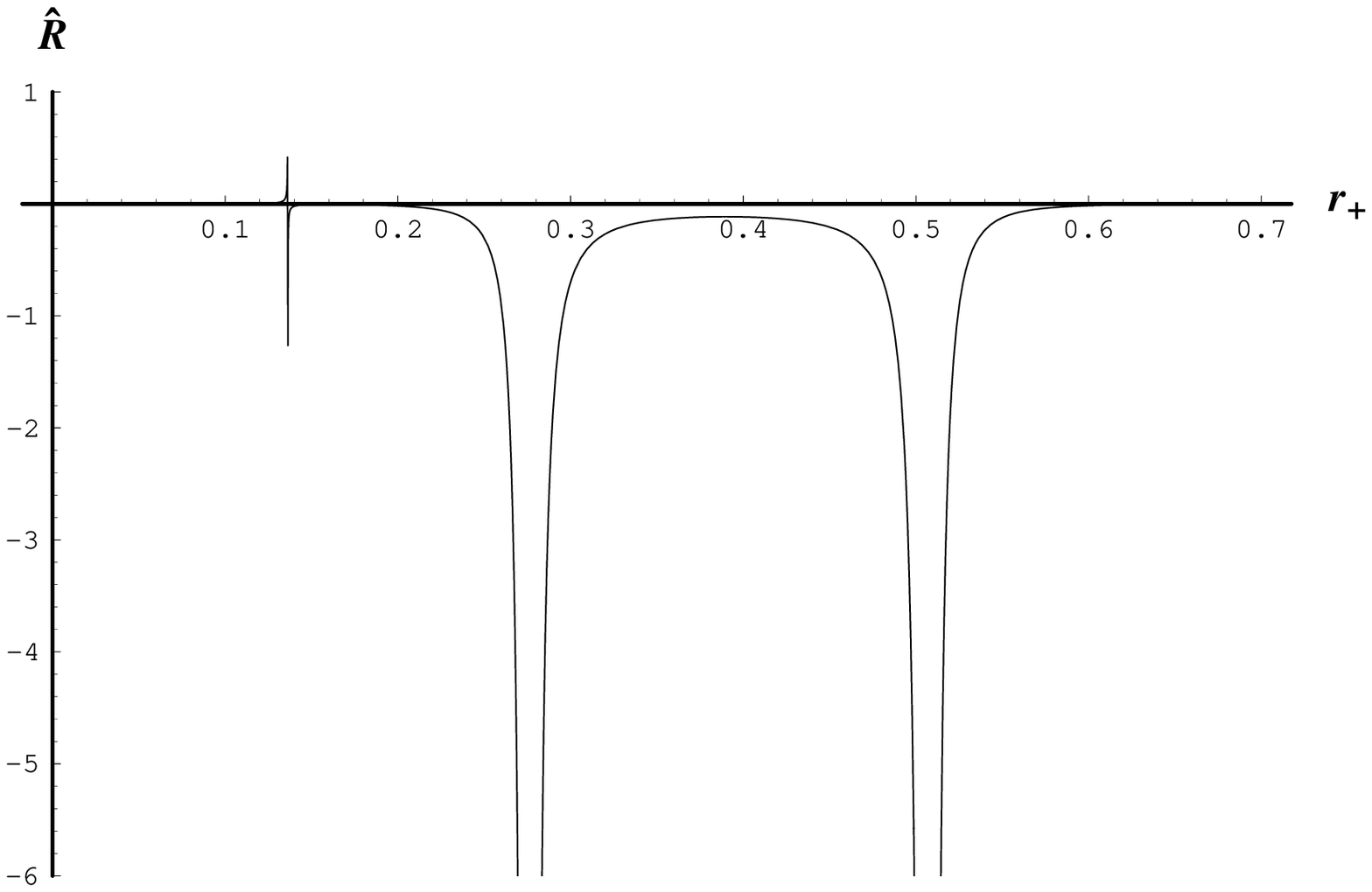}}
\caption{The thermodynamic scalar curvature $\hat R$ of the RN-AdS
black hole vs. the outer event horizon for fixed electric charge
$q<q_c$.}
\label{f2a}
\end{figure}
\begin{figure}
\resizebox{0.5\linewidth}{!}{\includegraphics*{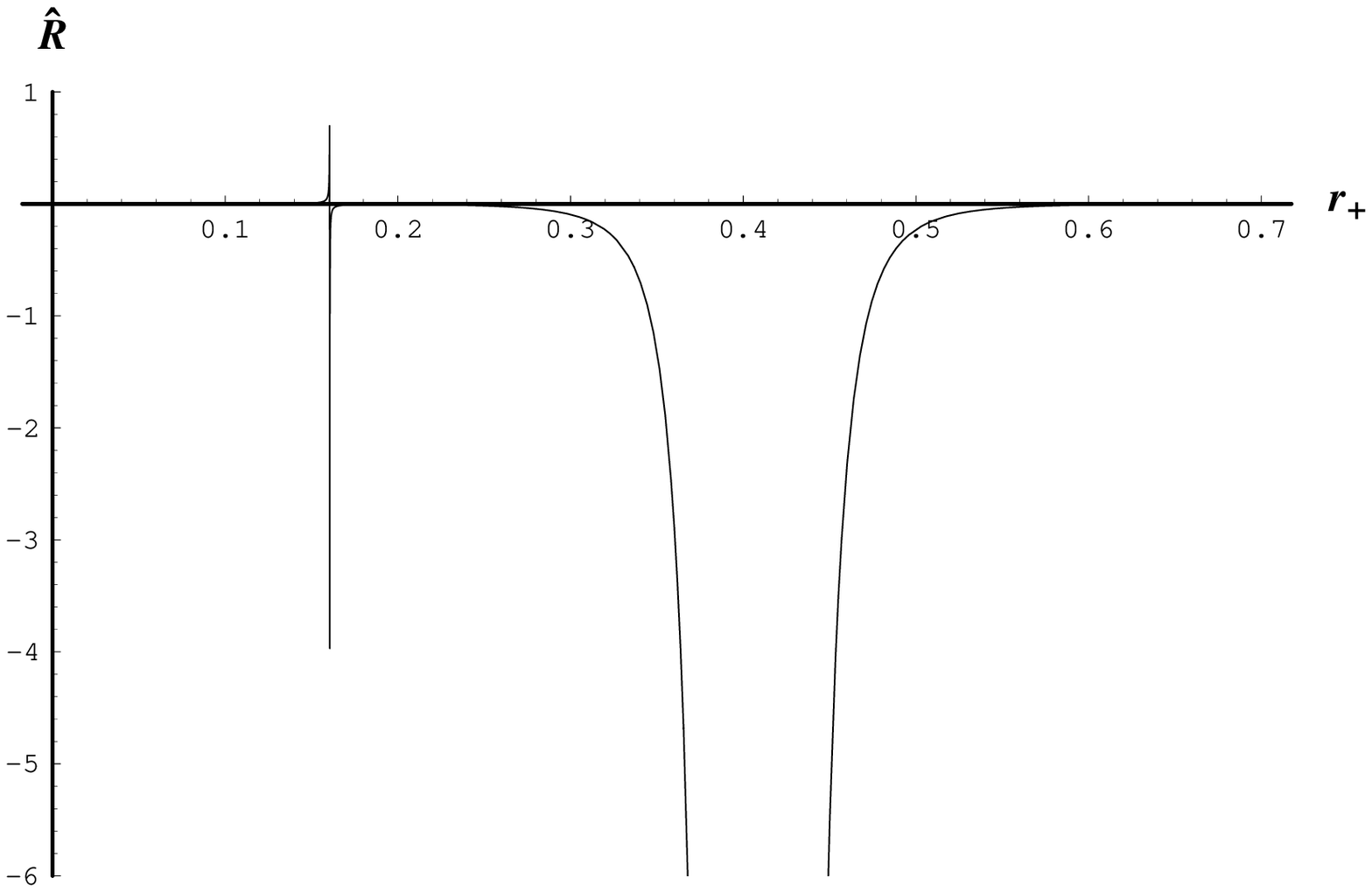}}
\caption{The thermodynamic scalar curvature $\hat R$ of the RN-AdS
black hole vs. the outer event horizon for fixed electric charge
$q=q_c$.}
\label{f2b}
\end{figure}
\begin{figure}
\resizebox{0.5\linewidth}{!}{\includegraphics*{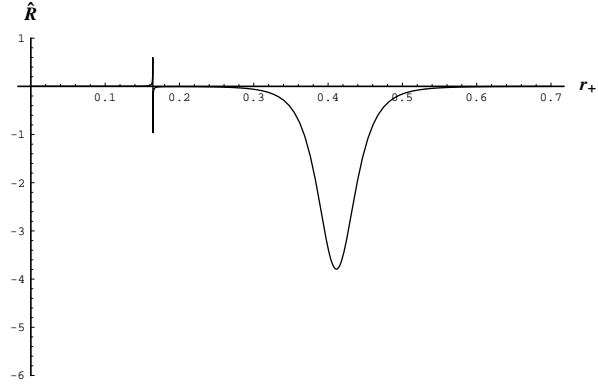}}
\caption{The thermodynamic scalar curvature $\hat R$ of the RN-AdS
black hole vs. the outer event horizon for fixed electric charge
$q>q_c$.}
\label{f2c}
\end{figure}

The thermodynamics and phase structure of the RN-AdS black holes
have been investigated in details in \cite{Cham1,Cham2,wu}. All
results are consistent with the expectation from the dual field
theory via holography. We will not reproduce those results here,
instead we just stress that Fig.\ref{f1} suggests a correspondence
of variables with $(\phi, q) \to (V, P)$ so as to establish the
analogy of RN-AdS black hole with the van der Waals model.
 As the cases of RN and Kerr black holes, the internal energy
of the black hole system should reflect the intrinsic properties
of the spacetime. Entropy should be a function of the internal
energy of system.  Therefore, the internal energy of the RN-AdS
black hole is
\begin{equation} \label{e33}
u = M  - \phi q
\end{equation}
and the first law of thermodynamics is
\begin{equation} \label{e34}
du = T dS  - q d\phi.
\end{equation}
Through the definition of the metric of thermodynamic metric
Eq.(\ref{e9}), we can obtain the scalar curvature of the
thermodynamic geometry
\begin{equation} \label{e35}
\hat R \propto -\{ (r_ +   - r_ -  )(l^2  + 3r_ + ^2  + 2r_ +  r_ -
+ r_ - ^2 )(3r_ + ^4  - l^2 r_ +  ^2  + 3l^2 q^2 )^2 \} ^{ - 1},
\end{equation}
which is rather complicated and listed in detail in the Appendix.
Fig.\ref{f2a}-\ref{f2c} show the behavior of $\hat R$ for the
various fixed electrical charge $q$. The very narrow vertical line
is caused by the singularity of the extremal black hole, whose
temperature is zero and this divergence reflects the breaking down
of thermodynamic description for the extremal black
holes~\cite{Ferr}. Of interest is the case with non-vanishing
Hawking temperature. We see from Fig.~\ref{f2a} that  when
$q<q_c$, the scalar curvature $\hat R$ has two negative infinities
where  $C_q$ diverges and $dq / d\phi =0$.  A second order phase
transition occurs when these two infinities merge into one in the
case of $q=q_c$ at $r_a = r_b =r_c$ in Fig.\ref{f2b}, which
indicates the appearance of a long range correlation in the
system. If $q>q_c$ as shown in Fig.\ref{f2c}, there is no singular
point for $\hat R$ except the extremal state of the black hole.

In order to see the thermodynamic behavior near the critical
point, the critical exponents can be introduced as
\begin{eqnarray}
\nonumber
 && (1)\quad q - q_c  \sim \left| {\phi  - \phi _c } \right|^\delta
\quad (T = T_c ),
\\ \nonumber
&& (2)\quad \phi  - \phi _c  \sim \left| {T - T_c } \right|^\beta
\quad (q = q_c ),
\\ \nonumber
&& (3)\quad C_q  \sim \left| {T - T_c } \right|^{ - \alpha } \quad
(q = q_c ),
\\
&& (4)\quad \kappa _T  \sim \left| {T - T_c } \right|^{ - \gamma }
\quad (q = q_c ).
 \end{eqnarray}
 According to the Ref.\cite{wu}, $\delta=3$,
$\beta=1/3$, $\alpha=2/3$ and $\gamma =2/3$ for the RN-AdS black
hole and they obey the scaling symmetry like  ordinary
thermodynamic systems
\begin{eqnarray}
\nonumber
 && \alpha  + 2\beta  + \gamma  = 2, \\ \nonumber
  && \alpha + \beta (\delta  + 1) = 2, \\ \nonumber
 &&  \gamma (\delta  + 1) = (2 - \alpha )(\delta  - 1), \\
 && \gamma  = \beta (\delta  - 1).
 \end{eqnarray}
On the other side, the thermodynamic scalar curvature $\hat R$ is
proportional to $|T-T_c|^{\alpha -2}$ near the critical point.
Ruppeiner~\cite{Rupp1} pointed out that the multiplication $R C_p
t^2$ should be a universal constant related to the critical
exponents $\beta$ and $\delta$~\cite{Rupp1}, that is
\begin{equation} \label{e36}
\ R C_p t^2  =  - \beta (\delta  - 1)(\beta \delta - 1)k_B,
\end{equation}
where $t = |T-T_c|/T_c$. The value of the r.h.s of Eq.(\ref{e36})
is zero if $\delta=3$ and $\beta=1/3$ are taken. In the case of
the RN-AdS black hole, our calculation shows that $R C_q t^2$ is
still a non-zero constant, which implies the difference between
thermodynamics of an ordinary system and of a black hole. This is
not surprising because there are some differences between black
holes and ordinary thermodynamic systems after all. For instance,
black holes can have negative heat capacity, entropy of black
holes is not an extensive quantity and so on.

\section{Conclusion And Discussion}

The Ruppeiner metric, defined as the Hessian matrix of entropy
with respect to internal energy and other extensive variables of a
thermodynamic system,  is closely related to the fluctuation
theory of equilibrium thermodynamics. It was argued that the
Riemannian geometry of the Ruppeiner metric can give insights into
the underlying statical mechanical system. In particular, it was
shown that the scalar curvature of the Ruppeiner geometry carries
much information on the phase structure of the thermodynamic
system and it diverges at critical points. Therefore it is quite
interesting to apply the geometry approach to black hole
thermodynamics.

The authors of  Refs.~\cite{Aman1,Aman2} defined the Ruppeiner
metric of RN (Kerr) black hole as the second derivatives of black
hole entropy with respect to the black hole mass and electric
charge (angular momentum). It was found that the Ruppeiner
geometry is flat and the scalar curvature vanishes for the RN
black holes, while it is curved and scalar curvature diverges at
the extremal limit for the Kerr black holes. Clearly it is easy to
see that a statistical model without any interaction cannot
re-produce thermodynamic properties of the RN black hole.
Furthermore, let us notice that for the RN and Kerr black holes,
their thermodynamic properties are quite similar. So it is not
easy to understand the results of \cite{Aman1,Aman2}. In this
paper, we have re-investigated the Ruppeiner geometry for the RN,
Kerr and RN-AdS black holes, based on the observations for an
analogy between the RN-AdS black hole and the Van der
Waals-Maxwell liquid-gas system. According to the analogy, an
interesting correspondence is $(q,\phi) \to (P,V)$, as we
explained above (also see \cite{Cham1,Cham2,wu} for details),
although it looks strange that the electric potential $\phi$ plays
the role of the extensive volume $V$.  Based on this observation,
we are enforced to think that the black hole mass does not stand
for the internal energy as in an ordinary thermodynamic system. As
a thermodynamic system, the internal energy of a black hole should
be the difference between the mass of black hole and the energy of
electric field for the RN black hole (the kinetic energy due to
rotation for the Kerr black hole), see (\ref{e7}) and (\ref{e18}),
respectively. Considering the entropy of black hole as a function
of internal energy and electric potential for charged black holes
(angular velocity for rotating black holes), we have calculated
the scalar curvature of resulting Ruppeiner metric for the RN,
Kerr and RN-AdS black holes.  We have found that the Ruppeiner
geometry is curved and the scalar curvature goes to negative
infinity at the phase transition points of Davies for the RN and
Kerr black holes, where heat capacity with a fixed charge (angular
momentum) diverges.
 In the case of non-extremal RN-AdS black hole, the
thermodynamic scalar curvature has two singularities when the
electrical charge $q$ is below its critical value $q_c$, which
indicates thermodynamic instability and the first order phase
transition as in Van de Waals liquid-gas system. The mergence of
these two singularities implies the existence of the second order
phase transition of the black hole at $q = q_c$. The black hole
behaves like a gas system with interaction for $q > q_c$. Our
results are completely consistent with those in
Refs.~\cite{Cham1,Cham2,wu}.

We have also investigated the critical behavior of the scalar
curvature $\hat R$ and found  that the scaling symmetry $RC_q t^2$
is still a non-zero constant, not as that of the ordinary
thermodynamic system predicated in Eq.(\ref{e36}). These results
indicate that while the Ruppeiner's theory does work well and is
consistent with the classical thermodynamics even in the system of
black holes, there are some differences between the thermodynamics
of black holes and that of ordinary thermodynamic system.  For
example, as it was noticed in \cite{wu} that critical exponents do
not match each other for the RN-AdS black hole and the van der
Waals model, although their phase structures are quite similar. In
addition, although the divergence of scalar curvature of the
Ruppeiner geometry is in good agreement with the black hole phase
transition in our setup, it is clearly needed to further
understand the physical meanings of the Ruppeiner geometry in the
black hole thermodynamics.

\section{Appendix}

In this appendix we give the Ruppeiner metric and its scalar
curvature for the RN-AdS black holes. We calculate the
thermodynamic metric as the function of inner and outer horizons.
Hence, the Jacobi matrix must be used here
\begin{equation} \label{a1}
J = \left( {\begin{array}{*{20}c}
   {r_{ + ,u} } & {r_{ + ,\phi } }  \\
   {r_{ - ,u} } & {r_{ - ,\phi } }  \\
\end{array}} \right) = \left( {\begin{array}{*{20}c}
   {u_{, r_ +  } } & {u_{, r_ -  } }  \\
   {\phi _{, r_ +  } } & {\phi _{, r_ -  } }  \\
\end{array}} \right)^{ - 1}.
\end{equation}
Using the definition of the thermodynamic metric Eq.(\ref{e9}) and
Jacobi Eq.(\ref{a1}), we obtain
\[
\hat g = \left( {\begin{array}{*{20}c}
   {S_{,uu} } & {S_{,u\phi } }  \\
   {S_{,\phi u} } & {S_{,\phi \phi } }  \\
\end{array}} \right),
\]
where
\begin{eqnarray}
\label{a2} \nonumber
 && S_{,uu}  = \frac{{2\pi l^2 r_ +  (3r_ +  ^4  - l^2 r_ +  ^2  + q^2
l^2 )}}{{(r_ +   - r_ -  )^3 (l^2  + 3r_ + ^2  + 2r_ +  r_ -   +
r_ - ^2 )^3 }},
\\ \nonumber
&& S_{,u \phi}=S_{,\phi u} = \frac{{4\pi l^2 r_ +  q(l^2 r_ +  ^2
- q^2 l^2 )}}{{(r_ + - r_ - )^3 (l^2  + 3r_ + ^2  + 2r_ +  r_ -
+ r_ - ^2 )^3 }},
    \\
 && S_{,\phi \phi }  = \frac{{\pi l^2 r_ +  ^3 z(r_ +  ,r_ -  )}}{{(r_ +
- r_ -  )^3 (l^2  + 3r_ + ^2  + 2r_ +  r_ -   + r_ - ^2 )^3 }},
\end{eqnarray}
and
\begin{eqnarray} \label{a5}
z(r_ +  ,r_ -  ) &=& 9r_ + ^6  + r_ + ^4 (6l^2  - 5r_ - ^2 ) - 5r_ -
^2 (l^2  + r_ - ^2 )^2  + 2r_ + ^3 (2l^2 r_ -   - 5r_ - ^3 ) \nonumber \\
&&+ r_ + ^2 (l^4  - 6l^2 r_ - ^2  - 15r_ - ^4 ) + 2r_ +  (2l^4 r_ -
- 3l^2 r_ - ^3  - 5r_ - ^5 ).
\end{eqnarray}
After a tedious calculation, finally we have the scalar curvature
\begin{equation} \label{a6}
\hat R =  - \frac{{B(r_ +  ,r_ -  )}}{{\pi (r_ +   - r_ -  )(l^2  +
3r_ + ^2  + 2r_ +  r_ -   + r_ - ^2 )(3r_ +  ^4  - l^2 r_ +  ^2  +
3q^2 l^2 )^2 }},
\end{equation}
where
\begin{eqnarray} \label{a7}
\nonumber
 B(r_ +  ,r_ -  ) &=& 54r_ +  ^7 r_ -   + l^2 r_ - ^2
(l^2 + r_ - ^2 )^2  + r_ + ^6 ( - 18l^2  + 84r_ - ^2 ) + 3r_ + ^5
(9l^2 r_ -   + 38r_ - ^3 )
\\ \nonumber
&&+ r_ + ^4 (9l^4  + 34l^2
r_ -  ^2  + 90r_ - ^4 ) + r_ + ^3 ( - 29l^4 r{}_ -  + 35l^2 r_ -
^3  + 60r_ - ^5 ) \\ \nonumber &&+ r_ +  ( - 2l^6 r_ - + 25l^2 r_
- ^5 ) + r_ + ^2 (l^6  + 30l^4 r_ - ^2 + 63l^2 r_ - ^4 + 30r_ - ^6
).
\end{eqnarray}
It is easy to check that $\hat R$ holds the form of Eq.(\ref{e10})
when the limit $l \to \infty$ is taken.

\begin{acknowledgments}

RGC thanks X.S. Chen and H.Lt Shi for useful discussions.  This
research was initiated when RGC was visiting the department of
physics, Fudan university, whose hospitality extends to him is
grateful. This work was supported in part by NNSF of China, by the
National Basic Research Program 2003CB716300 and the Foundation of
Education of Ministry of China. B. Wang's work was also supported
in part by Shanghai Science and Technology Commission.

\end{acknowledgments}


\end{document}